\documentclass{emulateapj}
\usepackage{color}
\usepackage{multirow}
\slugcomment{\today}

\def\kmsmpc{\rm km~s^{-1}~Mpc^{-1}}
\def\kms{\rm km~s^{-1}}
\def\age{\rm age}
\def\form{\rm form}
\def\Univ{\rm U}

\shorttitle{The age-redshift relation For LRGs}
\shortauthors{Liu et al.}

\begin{document}

\title{The Age-Redshift Relation For Luminous Red Galaxies obtained
from the Full Spectrum Fitting and its cosmological implications}

\author{Gaochao Liu\altaffilmark{1,2,3}, Youjun Lu\altaffilmark{1}, 
Xuelei Chen\altaffilmark{1,4}, Yongheng Zhao\altaffilmark{1}, 
Wei Du\altaffilmark{1,2}, and Xianmin Meng\altaffilmark{1}} 

\altaffiltext{1}{Key Laboratory of Optical Astronomy, National Astronomical 
Observatories, Chinese Academy of Science, Beijing 100012, China} 
\altaffiltext{2}{Graduate School of the Chinese Academy of Science, 
BeiJing 100049, China} 
\altaffiltext{3}{College of Science, China Three Gorges University, YiChang 
443002, China}
\altaffiltext{4}{Center of High Energy Physics, Peking University, Beijing 
100871, China}

\begin{abstract}

The relative age of galaxies at different redshifts can be used to infer the
Hubble parameter and put constraints on cosmological models. We select luminous
red galaxies (LRGs) from the SDSS DR7 and then cross-match it with the MPA/JHU
catalogue of galaxies to obtain a large sample of quiescent LRGs at redshift
$z\sim 0.03-0.39$.  The total 23,883 quiescent LRGs are divided into four
sub-samples according to their velocity dispersions and each sub-sample is
further divided into 12 redshift bins. The spectra of the LRGs in each redshift
and velocity bin are co-added in order to obtain a combined spectrum with
relatively high $S/N$.  Adopting the GalexEV/SteLib model, we estimate the mean
ages of the LRGs from these combined spectra by the full-spectrum fitting
method. We check the reliability of the estimated age by using Monte-Carlo
simulations and find that the estimates are robust and reliable. Assuming that
the LRGs in each sub-sample and each redshift bin were on average formed at the
same time, the Hubble parameter at the present time $H_0$ is estimated from the
age--redshift relation obtained for each sub-sample, which is compatible with
the $H_0$ value measured by other methods. We demonstrate that a systematic
bias (up to $\sim 20\%$) may be introduced to the $H_0$ estimation because of
recent star formation in the LRGs due to the later major mergers at $z\la 0.4$,
but this bias may be negligible for those sub-samples with large velocity
dispersions. Using the age--redshift relations obtained from the sub-sample
with the largest velocity dispersion or the two sub-samples with high velocity
dispersions, we find $H_0= 65^{+7}_{-3}\kmsmpc$ or $H_0= 74^{+5}_{-4}\kmsmpc$ 
by assuming a spatially flat $\Lambda$CDM cosmology. 
With upcoming surveys, such as the Baryon Oscillation
Spectroscopic Survey (BOSS),  even larger samples of quiescent massive LRGs may
be obtained, and thus the Hubble parameter can be measured with high accuracy
through the age--redshift relation.

\end{abstract}


\section{Introduction}

The expansion history of the universe are presently studied with a few
observational probes, such as  the supernova Ia, baryon acoustic oscillations
(BAO), weak gravitational lensing, and galaxy clusters, etc.  Each of these
probes has its pros and cons, and suffer from different systematic
uncertainties \citep[e.g.,][]{Fre10}.  A new observational probe of the cosmic
expansion history would be invaluable, and can provide additional cross check
with the results obtained from the existing methods. Combining the results
obtained by different means may further help to constrain robustly the
dynamical nature of the universe.

\citet{Jim02} proposed a novel approach to explore the expansion history of the
universe, which is based on the age--redshift relation of passively evolving
massive galaxies. Assuming that the passively evolving galaxies at different
redshifts were born approximately at the same time, the age of these galaxies
can then be taken as a cosmic chronometer. If the ages of such galaxies can be
accurately estimated, then this age--redshift relation may be used to determine
the cosmic expansion history.  Even if there is some systematic errors in the
absolute age measurements, it is argued that such errors could be canceled in
the relative age of these galaxies at different redshifts, thus providing a
good measurement of $H(z)$:
\begin{equation}
H(z) = -\frac{1}{1+z}\frac{dz}{dt}.
\label{eq:dtdz}
\end{equation}
Indeed, observations show that the most massive galaxies are mainly composed of
old stellar populations formed at redshifts $z > 1-2$, less than $1\%$ of their
present stellar mass is formed at $z<1$ \citep{Dun96,Spi97,Cow99,
Hea04,Tho05,Cim08,Tho10}, hence these galaxies are suitable for this
application.

\citet{Jim03} applied this method to a sample of massive galaxies at low
redshift by fitting their spectra with the single stellar population (SSP)
spectra based on the SPEED model \citep{Jim04}, and obtained $H_0=69
\pm~12\kmsmpc$.  \citet{Sim05} assembled a high redshift data set obtained from
the Gemini Deep Survey (GDDS) and some other archival data, and applied the
same method to estimate $H(z)$ for a large redshift range ($z\sim 0.1-1.8$).
These earlier works on the age--redshift relation adopted the SSP to fit each
galaxy spectrum in the sample, and selected the age of the oldest one in each
redshift bin as the envelop of the age. However, the poor signal-to-noise ratio
($S/N$) spectra of individual galaxies and the contamination from the telluric
emission and absorption may lead to uncertainties in the age estimates, and the
method of the oldest galaxy envelop draw results from a small number of
galaxies at the extremes of the distribution, which may also undermine the
validity of the result, and makes the method hard to use. 

To overcome this problem, \citet{Car10} obtained the combined spectra for those
luminous red galaxies (LRGs) with similar physical properties in each redshift
bin by co-adding their spectra, of which the $S/N$ is much higher than
individual galaxies. They then estimated the age of the combined spectra by
using the standard Lick absorption line indices, which may be regarded as the
mean age of a large sample of galaxies. They obtained the age--redshift
relation, but they did not use this relation to further constrain the Hubble
parameter. 

In this paper, we first select a LRG sample from the SDSS data release 7 (DR7).
In order to improve the $S/N$ and remove the contamination, we also use the
combined spectrum rather than the single spectrum of each galaxy.  However, we
adopt the full spectrum fitting method, different from the standard Lick
absorption line indices adopted by \citet{Car10}, to estimate the mean age of
the combined spectrum, and then obtain the age--redshift relation. Furthermore,
we also use the age-redshift relation obtained from the combined spectra to
constrain the Hubble parameter at the present time $H_0$ and analyze the possible
systematic bias in the estimated $H_0$.  The paper is organized as follows. In
Section~2, we describe the selection criteria of the LRG sample. In Section~3,
we provide the details of the fitting method and the age--redshift relation
estimated from the LRG sample. In Section~4, we constrain the Hubble
parameter by using the obtained age-redshift relation. Discussions on the
resulted age--redshift relation and the possible associated systematic bias are
given in Section~5.  Conclusions are summarized in Section~6.

\section{Sample selection}\label{sect:Sample} 

In order to obtain the age--redshift relation and use it to measure the Hubble
parameter, it is necessary to first select a large sample of passively evolving
galaxies that contains the oldest populations with homogeneous physical
properties. The SDSS is currently the largest survey that provides hundreds of
millions of detected objects with accurate photometric and astrometric
calibrations, and part of the objects have excellent spectra \citep{Pie03,
Hog01}. It is generally accepted that the LRGs are  passively evolving galaxies
and that they host the oldest stellar populations. Therefore, we pick our sample from
the LRGs of the SDSS DR7 \citep{Yor00,Aba09}.

For our purpose, it is necessary to determine the physical properties of the
LRGs with relatively high accuracy by using their spectra.  Considering that
some physical parameters such as the velocity dispersions and emission lines
are available in the MPA/JHU sample only for CUT I LRGs \citep[see][]{Eis01},
we select only from the CUT I LRGs. To obtain accurate estimates of the age
through the full spectrum fitting method, the selected galaxies should also
have sufficiently high $S/N$. Our selection criteria for LRGs are similar to
that described in \citet{Car10}, but with an additional restriction on the
$S/N$ as follows: 

\begin{itemize} 

\item selecting galaxies from Catalog Archive Server (CAS) database using the
\texttt{TARGET\_GALAXY\_RED} flag;

\item selecting galaxies with $S/N > 10$ per pixel (in the continuum of the
r-band wavelength range);

\item selecting galaxies which further satisfying the restrictions: 
       \texttt{specClass EQ `SPEC\_GALAXY'},\texttt{zStat EQ `XCORR\_HIC'},
       \texttt{zWarning EQ 0} , \texttt{eClass < 0},
       \texttt{z < 0.4} and \texttt{fracDev\_r > 0.8} 
       \citep[see more details in][]{Car10}.

\end{itemize}

\begin{figure}
  \centering
  \includegraphics[width=8.0cm, angle=0]{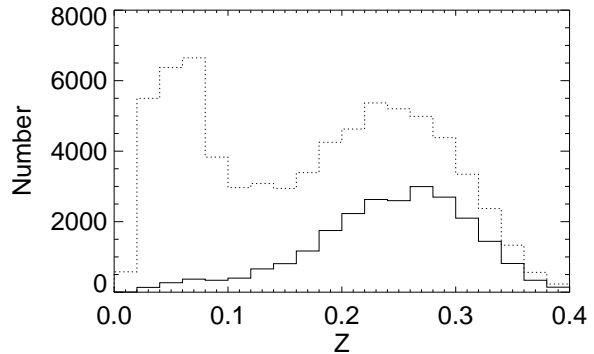}
  \caption{The redshift distribution of LRGs. The solid histogram represents 
           the distribution of LRGs in our final sample and the dotted 
           histogram represents the distribution of all LRGs in our paper 
           whose the lower-redshift peak is mostly contributed by the spiral 
           bulges. 
          }
   \label{fig:f1} 
\end{figure}

According to the criteria above, 71,971 LRGs are selected from the SDSS DR7 
and their redshift distribution is shown in Figure~\ref{fig:f1} (the dotted
histogram). These LRGs are probably contaminated by the bulges of late-type
galaxies at low redshift, due to the limited fiber size (3\arcsec) of the SDSS
spectrograph. Such late contaminants would however have new star formations,
and it is well known that the [O{\sc{ii}}] and $H_\alpha$ lines are indicators
of star formation, hence we can remove the spiral bulges from the selected
sample and obtain an sample of quiescent LRGs by using the spectral line data.
Here we use the MPA/JHU spectral line
data\footnote{http://www.mpa-garching.mpg.de/SDSS/DR7/raw\_data.html}, which
has been widely used in selecting quiescent galaxies in the literature. We
select those LRGs as quiescent only if their $H_\alpha$ and [O{\sc{ii}}] line
emission are consistent with zero at $2\sigma$ level. With this criterion, we
obtain 27,208 quiescent LRGs. We note here that using other emission lines to
select quiescent galaxies may result in a similar quiescent LRG sample, as
\citet{Car10} pointed out that all other emission lines (such as the joint
constraint of $H_\beta$ and [O{\sc{iii}}] or N{\sc{ii}} and S{\sc{ii}}) show
similar zero-emission line distributions.

To estimate the Hubble expansion rate $H(z)$ by using the age-redshift
relation, it is important to select samples of galaxies with homogeneous
physical properties as demonstrated by \citet{Cra10}. For this reason, the
quiescent LRGs selected above are divided into four sub-samples according to
their velocity dispersions, which are listed in the MPA/JHU galaxy catalog. The
velocity dispersion bins for each sub-sample of the LRGs are $200\kms-230\kms$,
$230\kms-260\kms$, $260\kms-290\kms$ and $290\kms-320\kms$, respectively. We
denote these four sub-samples as sub-sample I, sub-sample II, sub-sample III,
and sub-sample IV, respectively. In each of the velocity dispersion bins, the
number of galaxies is still sufficiently large for the following co-adding
spectra to reach a high $S/N$ ($> 40$). We do not consider galaxies with
velocity dispersions larger than $320\kms$, as the total number of those
galaxies at $z<0.14$ is too small ($<30$). In addition since the total number of 
galaxies with velocity dispersion less than $200\kms$ is 1843, much smaller than
 the number of galaxies in all other sub samples, we also exclude these galaxies.
 The number of galaxies in each sub-sample is listed in
Table~\ref{tab:t1}, and after excluding those LRGs with $\sigma_v>320\kms$ or
$\sigma_v<200\kms$, the total number of the final sample is 23,883. The
redshift distribution of these galaxies is shown in Figure~\ref{fig:f1} (the
solid histogram). 

\begin{deluxetable}{ccc}
\tabletypesize{\scriptsize}
  \tablecaption{The total number of galaxies in each sub-sample.}
  \tablewidth{0pt}
\tablehead{ \colhead{Sample} & \colhead{ Velocity Dispersion Range}
& \colhead{Number} } 
\startdata
sub-sample I   & $200\kms<\sigma_{v}\le230\kms$ & $4756$  \\
sub-sample II  & $230\kms<\sigma_{v}\le260\kms$ & $8748$  \\
sub-sample III & $260\kms<\sigma_{v}\le290\kms$ & $7149$  \\
sub-sample IV  & $290\kms<\sigma_{v}\le320\kms$ & $3230$ \\
total          & $200\kms<\sigma_{v}\le320\kms$ & $23,883$ 
\enddata
\label{tab:t1}
\end{deluxetable}

\section{Spectral fitting methods}\label{sect:full spectrum}

There are three commonly adopted methods for measuring the age and metallicity
of a stellar system from its spectrum: (1) the SED fitting, (2) the Lick
indices fitting, and (3) the full spectrum fitting. The first method is only
sensitive to the general shape of the continuum, the second one focuses on
using the strength or equivalent width of lines and specific spectrum features,
and the third one accounts all the information of the spectrum, including both
the continuum and the lines and specific features. The full spectrum fitting
method has several advantages, such as being insensitive to extinction or flux
calibration errors, and it is also not limited by the physical broadening of
lines since the internal kinematics is determined simultaneously with the
population parameters \citep{Kol08},
though it is insensitive to the element ratio effects 
because of yet no available models about these.
 On the other hand, it is more sensitive to
the wavelength range of the spectrum adopted in the fitting and the resolution
of the spectrum compared with the fitting with  the Lick indices.

\subsection{The \texttt{ULySS} software}\label{sect:ulyss}

\texttt{ULySS} is an open-source software package developed by a group in
Universit\'e de Lyon, which implements the full-spectrum fitting to study
physical properties of stellar populations. In \texttt{ULySS}, an observed
spectrum is fitted by a model spectrum, adopting a linear combination of
non-linear components, optionally convolved with a line-of-sight velocity
distribution (LOSVD) and multiplied by a polynomial function.  The
multiplicative polynomial is adopted to absorb errors of the flux calibration,
Galactic extinction and other factors which may affect the shape of the
spectrum.  It minimizes $\chi^2$ value by the MPFIT function when matching an
observed spectrum with the model ones. The line spread function (LSF), an
analogy to the point spread function (PSF) for images, is also introduced in
\texttt{ULySS} in order to effectively match the resolution of the model
spectrum to the observations. For details about \texttt{ULySS}, the readers are
referred to \citet{Kol09a}. Since the full spectrum fitting method may be
sensitive to the wavelength range of the spectrum adopted in the fitting, we
adopt the GalexEV/SteLib model, a popular single stellar populations (SSPs)
synthesis model which covers the largest wavelength range, i.e. $3200$\AA$-9500$\AA. This wavelength range includes the Ca{\sc{ii}} triplet
($\lambda\lambda 8498, 8542, 8662$\AA), which is a prominent feature
produced primarily by an old population of red-giants and thus important for
determining the age of the quiescent LRGs \citep{Dia89}.

The GalexEV/SteLib(GS) population model is produced by the isochrone synthesis
code of BC03 \citep{Bru03}, which is widely used in SDSS data analysis. It use
the SteLib library which contains 249 spectra, but only 187 stars have measured
metallicity and can be used to compute the predicted spectra with a 3\AA
 spectral resolution. The GS model adopts the Padova~94 isochrones \citep{Ber94}
and the Chabrier IMF \citep{Cha03}. Totally 696 SSPs, covering the age of $0.1
\sim 20$~Gyr and the [Fe/H] of $-2.3\sim 0.4$~dex, are included in the GS
model. The relevant information of the GS model is given in Table~\ref{tab:t2}.

\begin{deluxetable*}{llccccll}
\tabletypesize{\scriptsize}
\tablecaption{The settings in the GS model. }
\tablewidth{0pt} \tablehead{ \colhead{Model} & \colhead{Library} & \colhead{Resolution} 
& \colhead{Wavelength} & \colhead{Age} & \colhead{Z} & \colhead{IMF} & \colhead{Track} \\
& & \colhead{(\AA)} & \colhead{(\AA)} & \colhead{(Gyr)} & 
\colhead{(dex)} & & 
}
\startdata
GS & SteLib & 3 & 3200-9500 & 0.1-20 & $-2.3$-0.4 & Chabrier & Padova~94 
\enddata
\label{tab:t2}
\end{deluxetable*}

\subsection{Model pre-treatment}\label{sect:spectrum pre-treatment}

The resolution match is a key issue in the model fitting because the resolution
of a model spectrum is usually different from that of the observational data.
It is necessary to transform either the model spectrum or the observed spectrum
to match the resolution of the other one.  The spectral resolution is
characterized by the instrumental broadening or the LSF. In practice, the LSF
is not necessarily a Gaussian and may vary with wavelength (see more details on
the LSF in \citet{Kol08,Kol09a}). In \texttt{ULySS}, three types of
calibrations (arc lamp, standard star, twilight spectrum) can be used to
determine the relative LSF between the model and the observation. In this
paper, we use the standard star to do the calibrations. \texttt{ULySS} adopts a
convolution of the model with a series of LSFs, which are determined at some
wavelengths, and then interpolates linearly in the wavelength range between two
convolved models.

We follow the steps described in \citet{Kol09a} and \citet{Du10} to match the
resolution of the spectra for galaxies in our sample with that from the GS
model. We use the spectrum of SDSS standard star, which is already contained in
the \texttt{ULySS} package, as a representative of the SDSS observations. For
the GS model spectrum, the relative LSF is obtained in \texttt{ULySS} by
comparing the spectrum of the observed spectrum of the SDSS standard stars with
the GS model spectrum of the stars with the same physical properties. We then
adopt this relative LSF to generate the resolution-matched GS model spectrum by
the LSF convolution function in the \texttt{ULySS} package. Below when we
mention the GS model spectrum we are actually referring to such
resolution-matched ones.

\subsection{Spectrum fitting}\label{sect:SSP fitting} 

We adopt the SSPs given by \texttt{ULySS} to fit the combined spectrum or the
spectrum of each galaxy in our sample but do not consider the detailed star
formation history (SFH) of each galaxy, as the sample is almost homogeneous and
the galaxies in it are passively evolving. We defer the discussion of the
effect due to the difference in the SFH of each galaxy to
Section~\ref{sect:discussion}.

\subsubsection{Combined Spectrum}
\label{sect:combined spectrum} 

High $S/N$ spectra are essential to obtain accurate estimates of the age of
galaxies. We have tested the effect of the $S/N$ on the age uncertainty. We
found that the uncertainties in age estimates are about $20\%$ and $10\%$ for
galaxy spectra with $S/N=20$ and $40$, respectively. In order to apply the
age--redshift relation effectively, the uncertainties in age estimates need to
be smaller than $10\%$. For the majority of galaxies in our sample, however,
their SDSS spectra have $S/N \la 20$, which leads to  $>20\%$ uncertainties in
age estimates. To overcome this problem, we choose to co-add the spectra of
galaxies in each sub-sample (and each red-shift bin) to obtain a combined
spectrum with significantly higher $S/N$ ($>40$).  

To do this, we first correct the foreground Galactic extinction by using the
reddening maps of \citet{Sch98} for each galaxy.  As the number of galaxies at
$z<0.03$ is too small in all of the sub-samples, these galaxies are neglected
in the following analysis. The galaxies in each sub-sample are then divided
into $12$ redshift bins from $z=0.03$ to $z=0.39$ with a redshift step of
$\delta z=0.03$. A combined spectrum is then obtained for each redshift and
velocity dispersion bin by co-adding the rest frame spectra of all the galaxies
in that bin through the \texttt{IRAF} task \texttt{SCOMBINE}. We thus obtained
12 combined spectra over redshift bins from $0.03$ to $0.39$ for each
sub-sample.

\begin{figure*}[!htbp]
\centering
 \includegraphics[scale=0.6]{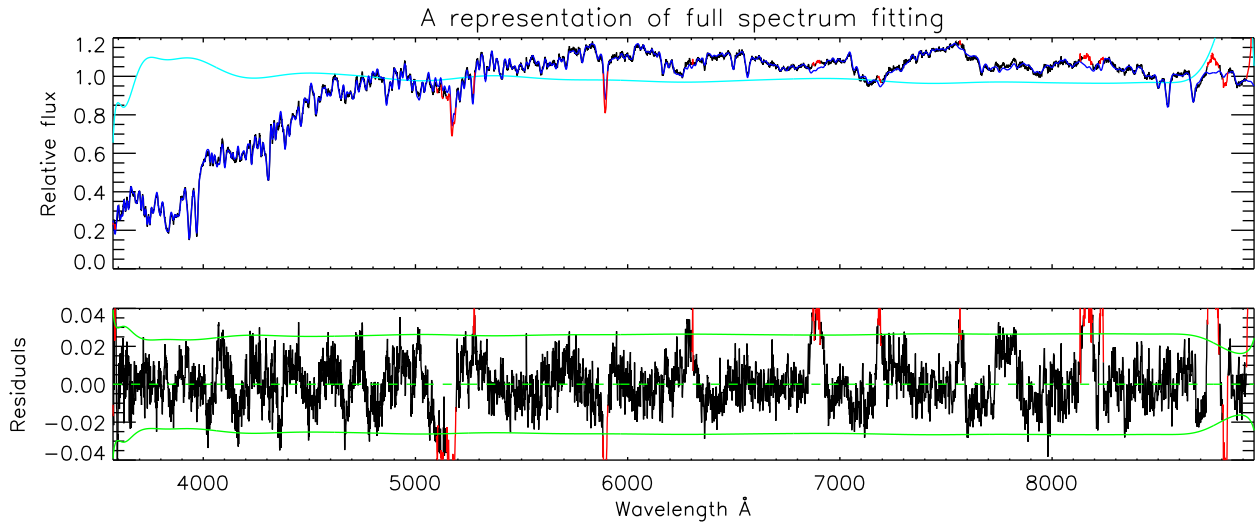}
 \caption{
The combined spectrum of the galaxies in the first redshift bin of the
sub-sample II and  its best-fit by adopting the GS model. In the top panel,
black and blue lines show the combined spectrum and its best-fit model
spectrum, respectively. These two spectra are almost superimposed on each other
and the black line can be seen only when zooming in the Figure. The cyan line
is the multiplicative polynomial to absorb the effects of an imprecise flux
calibration and of the Galactic extinction. The red regions are rejected from
the fitting (rejection of flagged telluric lines and automatic rejection of
outliers). The bottom panel shows the residuals of the best-fit. The solid
green lines mark the $1\sigma$ deviation, and the dashed green line represents
zero residuals.
}
\label{fig:f2}
\end{figure*}

The combined spectra are then fitted by \texttt{ULySS}.  \texttt{ULySS} use
Levenberg-Marquardt routine to search the parameter space to get the
minimization of $\chi^2$. So it needs some initial guess value to begin
searching the parameter space.  In this paper, we set the initial guess for
 the age to 8 Gyr and for metallicity to solar metallicity since the LRGs are
 expected to be old and metal rich.  We do not set any limit
on the allowed age and the abundance of metallicities of the model spectra.
The fitting is performed in the whole wavelength range covered by the combined
spectra.  The estimates on the age, metallicity and velocity dispersion of the
model galaxies and the associated errors are then obtained for the best-fit
model to the combined spectrum.  For illustration, Figure~\ref{fig:f2} shows
the combined spectrum in the first redshift bin of the sub-sample II, and the
best-fit model spectrum by the GS model and its residues. The fitting results
for the sub-sample II are listed in Table~\ref{tab:t3}.

 \begin{deluxetable*}{ccccc}
\tabletypesize{\scriptsize}
\tablecaption{Results of a SSP fitting with the GS model for sub-sample II.}
\tablewidth{0pt} \tablehead{
\colhead{Redshift Interval} &  \colhead{SSP-equivalent age}
 & \colhead{SSP-equivalent [Fe/H]} & \colhead{velocity dispersion}\\
\colhead{} & \colhead{(Gyr)}&\colhead{(dex)}
&\colhead{($\kms$)} } 
\startdata
$0.03 < z \leq 0.06$&7.13$\pm$0.53&0.17$\pm$0.01&243.5$\pm$4.0\\
$0.06 < z \leq 0.09$&6.52$\pm$0.43&0.17$\pm$0.01&239.2$\pm$3.7\\
$0.09 < z \leq 0.12$&6.31$\pm$0.40&0.16$\pm$0.01&236.0$\pm$3.7\\
$0.12 < z \leq 0.15$&6.01$\pm$0.40&0.17$\pm$0.01&239.0$\pm$3.7\\
$0.15 < z \leq 0.18$&6.11$\pm$0.45&0.15$\pm$0.01&240.7$\pm$3.9\\
$0.18 < z \leq 0.21$&5.49$\pm$0.24&0.18$\pm$0.01&242.4$\pm$3.8\\
$0.21 < z \leq 0.24$&5.37$\pm$0.22&0.17$\pm$0.01&243.0$\pm$3.5\\
$0.24 < z \leq 0.27$&4.98$\pm$0.22&0.15$\pm$0.01&242.7$\pm$3.5\\
$0.27 < z \leq 0.30$&4.96$\pm$0.19&0.14$\pm$0.01&247.6$\pm$3.3\\
$0.30 < z \leq 0.33$&4.78$\pm$0.16&0.12$\pm$0.01&247.4$\pm$3.4\\
$0.33 < z \leq 0.36$&3.82$\pm$0.24&0.16$\pm$0.02&243.7$\pm$3.5\\
$0.36 < z \leq 0.39$&3.64$\pm$0.19&0.17$\pm$0.02&248.9$\pm$3.2
\enddata
\tablecomments{Column 1 is the redshift interval, columns 2, 3 and 4 are
              the SSP-equivalent age in unit of Gyr, the [Fe/H] in unit 
              of dex, the velocity dispersion in unit of $\kms$ and their 
              associated errors, respectively.}   
\label{tab:t3}
\end{deluxetable*}

\begin{figure*}[!htbp] 
\centering
\includegraphics[scale=1.0,width=16.0cm]{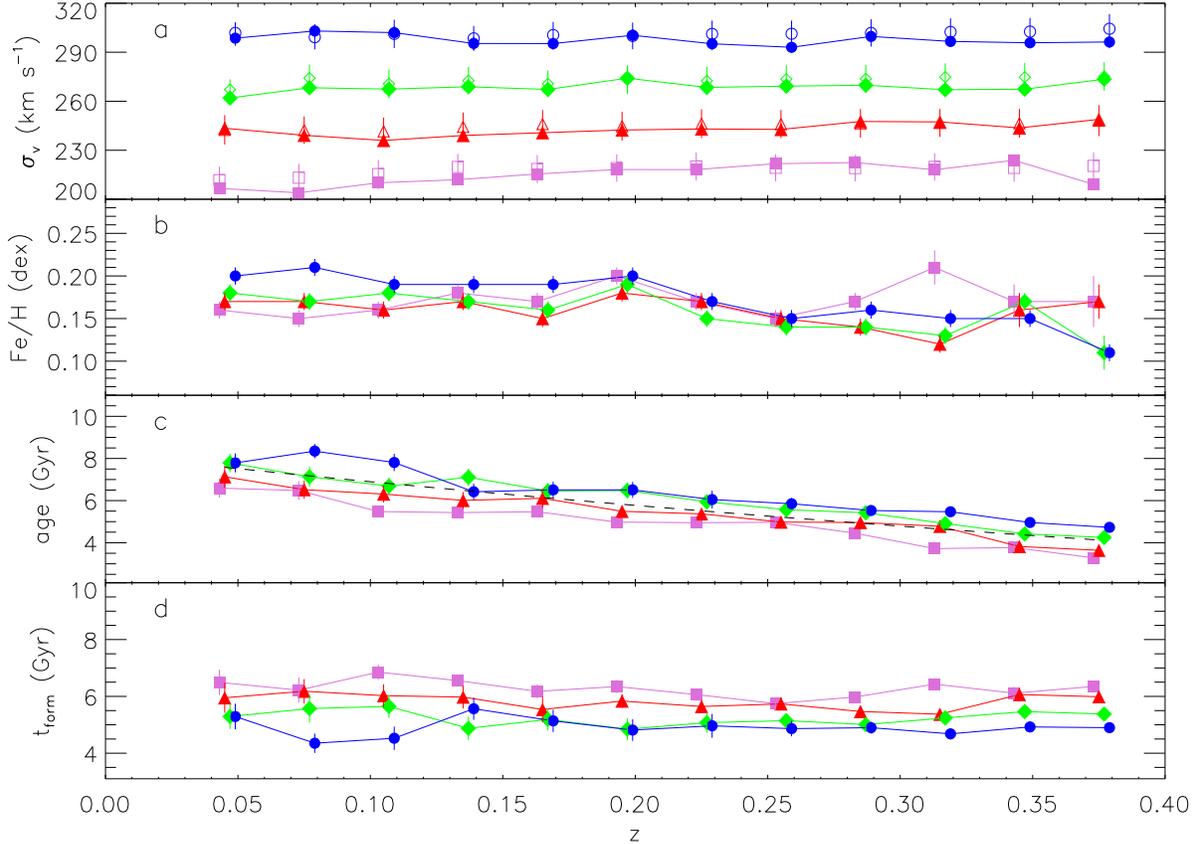} 
\caption{ 
The mean galactic properties extracted from the combined spectra by using the
GS model. Panels (a), (b), (c) and (d) show the velocity dispersions,
metallicities, mean luminosity-weighted ages and mean formation time of the
LRGs obtained through their combined spectra in each redshift bin for each of
the four sub-samples, $200\kms<\sigma_{v} \le230\kms$ (purple squares),
$230\kms <\sigma_{v} \le 260\kms$ (red triangles), $260\kms <\sigma_{v} \le
290\kms$ (green diamonds), and $290\kms <\sigma_{v} \le 320\kms$ (blue
circles), respectively. For clarity, those symbols for the same redshift bin
but different sub-sample are put offset by $\delta z=-0.002$, $0$, $0.002$, and $0.004$ 
along the horizontal axis for the four sub-samples with velocity
dispersions from low to high, respectively.  The galactic velocity dispersions
obtained directly from the MPA/JHU catalogue are shown as open squares,
triangles, diamonds, and circles for each sub-sample, respectively.  In the
panel (c), the dashed line indicates $t_{\Univ}(z)-5.5\,\textnormal{Gyr}$,
where $t_{\Univ}(z)$ is the age of the universe for a $\Lambda$CDM cosmology
and is for reference only. The concordant cosmological model, i.e.,
$H_0=71\kmsmpc$ and $\Omega_{\rm m}=0.27$, is adopted to obtained the mean 
galaxy formation time shown in panel (d).
} 
\label{fig:f3} 
\end{figure*}

Figure~\ref{fig:f3} illustrates the best fit result obtained by adopting the GS
model. The top, middle and bottom panel in the Figure shows the velocity
dispersions, metallicities and ages for each redshift bin in each of the four
sub-samples, respectively. As seen from the top panel of Figure~\ref{fig:f3}, the fitting results on
the velocity dispersions are consistent with the ones given in the MPA/JHU
catalogue, which are shown in Figure~\ref{fig:f3} as open squares, triangles,
 diamonds, and circles,  respectively.
The middle panel shows that the galaxies in each sub-sample have
similar metallicities, confirming that our samples are almost homogeneous,
though of course subtle differences remain. From the bottom panel, it is clear
that the mean age ($t_{\rm age}$) decreases with redshifts from $0.03$ to
$0.39$. Apparently, the age of the galaxies with higher velocity dispersion
tend to be somewhat older than those with lower velocity dispersion, which is
consistent with the well known ``downsizing'' formation of galaxies, i.e., the
bigger and more massive galaxies were formed earlier, while the small galaxies
formed later \citep{Cowie96}. 
Assuming that the $t_{\age}-\sigma_v$ relation follows a
power-law, i.e., $t_{\age} \propto \sigma_v^{\gamma}$, we fit the
relationship for each redshift bin and obtain the slope $\gamma$. The
mean value of $\gamma$ for all the redshift bins (and its standard
deviation) is $\simeq 0.77\pm 0.25$. We may also first obtain the mean
$t_{\age}$ for each sub-sample and then fit the mean $t_{\rm
age}-\sigma_v$ relationship and find $\gamma\simeq 0.79\pm 0.24$.

A number of studies have obtained the relationship between the age 
($t_{\age}$) and velocity dispersion ($\sigma_v$) of early-type 
galaxies or LRGs and shown that the $t_{\age}$ depends $\sigma_v$. For 
example, \citet{Cal03} analyzed the integrated spectra of 175 nearby
early-type galaxies by using higher order Balmer lines as the age
indicators and found that early-type galaxies with lower $\sigma_v$
have smaller luminosity-weighted mean ages. The data obtained by
\citet{Cal03} suggests that the slope of the $t_{\age}-\sigma_v$
relation is roughly $0.8-1.2$ \citep{Nel05}.  \citet{Tho05} studied
the spectra of 124 early-type galaxies in both high and low density
environments by using the absorption line indices, and they also found
that $t_{\age}$ correlates with $\sigma_v$. Adopting the data in
\citet{Tho05}, the slope of the $t_{age}-\sigma_v$ relation is found to
be $0.78\pm0.23$ \citep{Nel05}.  \citet{Nel05} investigated the
spectroscopic line strength of 4097 red-sequence galaxies in 93
low-redshift galaxy clusters, and they found $t_{\age}\propto 
\sigma_v^{0.59\pm0.13}$. \citet{Smi09} studied the spectra of 232
quiescent galaxies in the Shapley supercluster and found $t_{\age}
\propto \sigma_v^{0.40}$. As discussed in \citet{Nel05}, the differences
in sample selection criteria and emission treatment may be account for
the differences of $t_{\age}-\sigma$ scaling relation.  Considering of
these differences, our results are well consistent with those obtained
by previous works.

\subsubsection{Reliability of the fitting results}\label{sect:reliability}

It is important to check the reliability of the fitting results to ensure it
is not highly dependent on the particular synthesis technique adopted, or
affected by the existence of multiple solutions due to degeneracies among the
model parameters. 

\citet{Kol08} analyzed the spectra of Galactic clusters using \texttt{ULySS},
and they found that stellar populations of these clusters obtained from the
model are well consistent with that obtained from the color-magnitude diagrams.
\citet{Kol09b} further analyzed the detailed star formation history of 16 dwarf
galaxies by using either \texttt{ULySS} or \texttt{STECKMAP}, and they found
that the two programs give remarkably consistent results. In addition,
\citet{Mic07} adopted two different techniques, i.e., the  Lick/IDS index
system and the full spectrum fitting method, to test the reliability of the
estimates of the ages and metallicities of 16 dwarf elliptical galaxies, and
they found these two techniques give consistent results on the age and the
metallicity, with an rms error of 1.63 Gyr in age and 0.09 dex in [Z/H].
\citet{Du10} synthesized the star formation histories and evolution of 35
brightest E+A galaxies from the SDSS DR5, and demonstrated the robustness of
the \texttt{ULySS} technique in measuring the age and metallicity of stellar
systems.  These studies show that the fitting results with different synthesis
technique are consistent with each other, and the results of our \texttt{ULySS}
fitting should be robust.

\begin{figure*}[!htbp]
\centering
\includegraphics[scale=1.0,width=16.0cm]{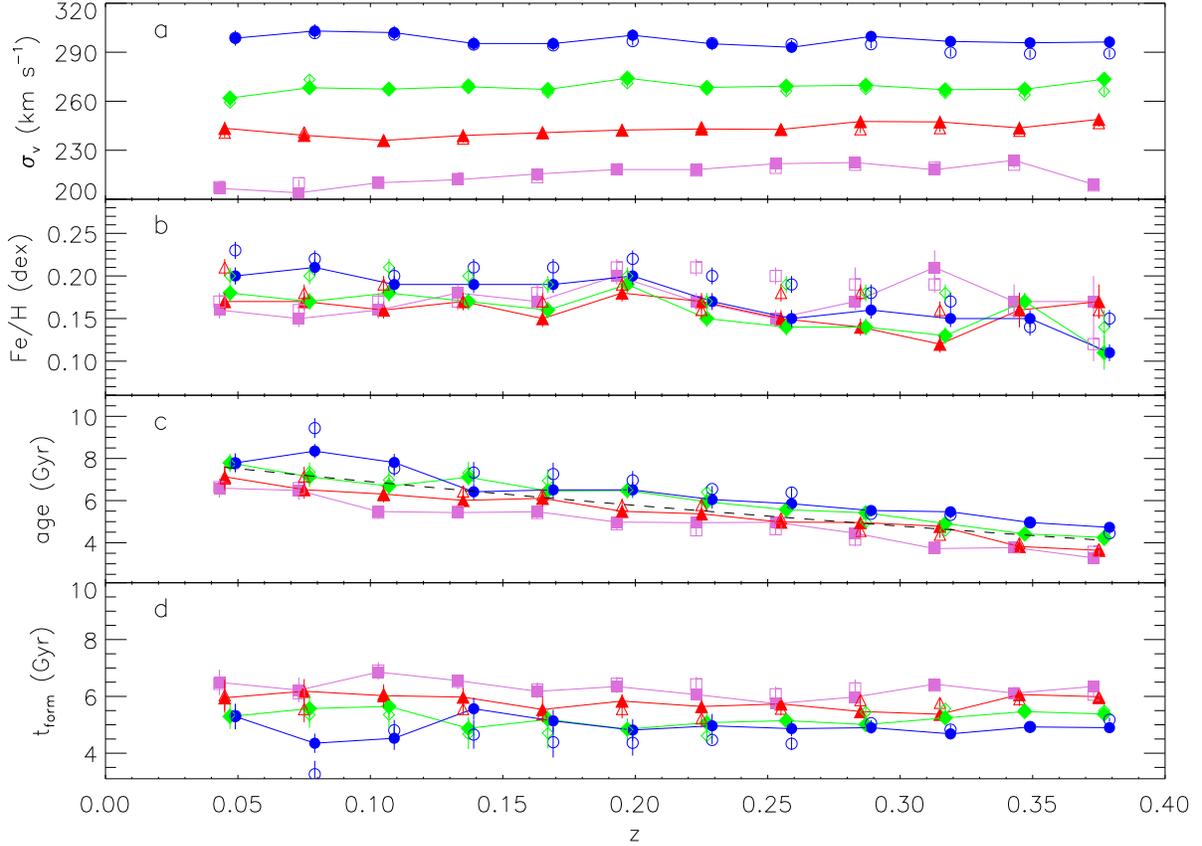}
\caption{Legends are similar to that of Figure~\ref{fig:f3}, except that the 
Monte-Carlo simulation results (open symbols) are additionally shown here.
And the Monte-Carlo simulations are performed to visualize the degeneracy
and validate the estimation of the errors (see Section~\ref{sect:reliability}).
}
\label{fig:f4} 
\end{figure*}

We perform Monte-Carlo simulations to visualize the degeneracies and validate
the errors by simulating the effect of the noise. For each combined spectrum,
we perform 1000 simulations.  In each of the simulations a random Gaussian
noise is added to the combined spectrum and the amplitude of the added noise is
set to the estimated noise associated to the combined spectrum. We then get the
mean values of the age, abundance of metallicity and velocity dispersion, and
their standard deviations, correspondingly.  Figure~\ref{fig:f4} shows the
results of the 1000 Monte-Carlo simulations (open symbols) and the original
best fit shown in Figure~\ref{fig:f3} (filled symbols). According to
Figure~\ref{fig:f4}, we conclude that in most of cases, the value of mean
velocity desperation and mean age for every bin of the 1000 Monte-Carlo
simulations lead to the best fitting solutions, though there are still some
deviations especially for sub-sample IV, which may be caused by relatively small
number of galaxies used to obtain the combined spectrum. Since we only need to
model the age--redshift relation, the diversity of metallicity between the
best fitting value and Monte-Carlo simulations value do not affect
our conclusion.

The general agreement of the above series tests suggest that the \texttt{ULySS}
technique is robust in determining the age, metallicity and velocity dispersion
of stellar systems, and the fitting results on the physical properties of
galaxies obtained from the full spectrum fitting are also secure.

\section{Constraints on the Hubble parameter}\label{sect:Hubble_parameter}

The age--redshift relation obtained from observations can be directly fitted by 
\begin{equation}
t_{\age}=t_{\Univ}-t_{\form}, 
\label{eq:tage}
\end{equation}
where $t_{\age}$ is the mean age obtained from the combined spectrum,
$t_{\form}$ is the mean formation time of the quiescent galaxies and assumed to
be a constant for each sub-sample, and $t_{\Univ}$ is the age of the universe.
According to the $\Lambda$CDM cosmology, $t_{\Univ}$ at redshift $z$ is given
by
\begin{equation}
t_{\Univ}= \frac{1}{H_0} \int_{z}^{\infty} \frac{dz}{(1+z) 
\sqrt{\Omega_{\rm m}(1+z)^3 + \Omega_{\Lambda}}},
\label{eq:tuniv}
\end{equation}
which depends not only on the value of $H_0$ but also the composition of the
universe, i.e., $\Omega_{\rm m}$ and $\Omega_{\Lambda}$. 

\begin{figure*} 
\centering
\includegraphics[scale=1.0,width=16.0cm]{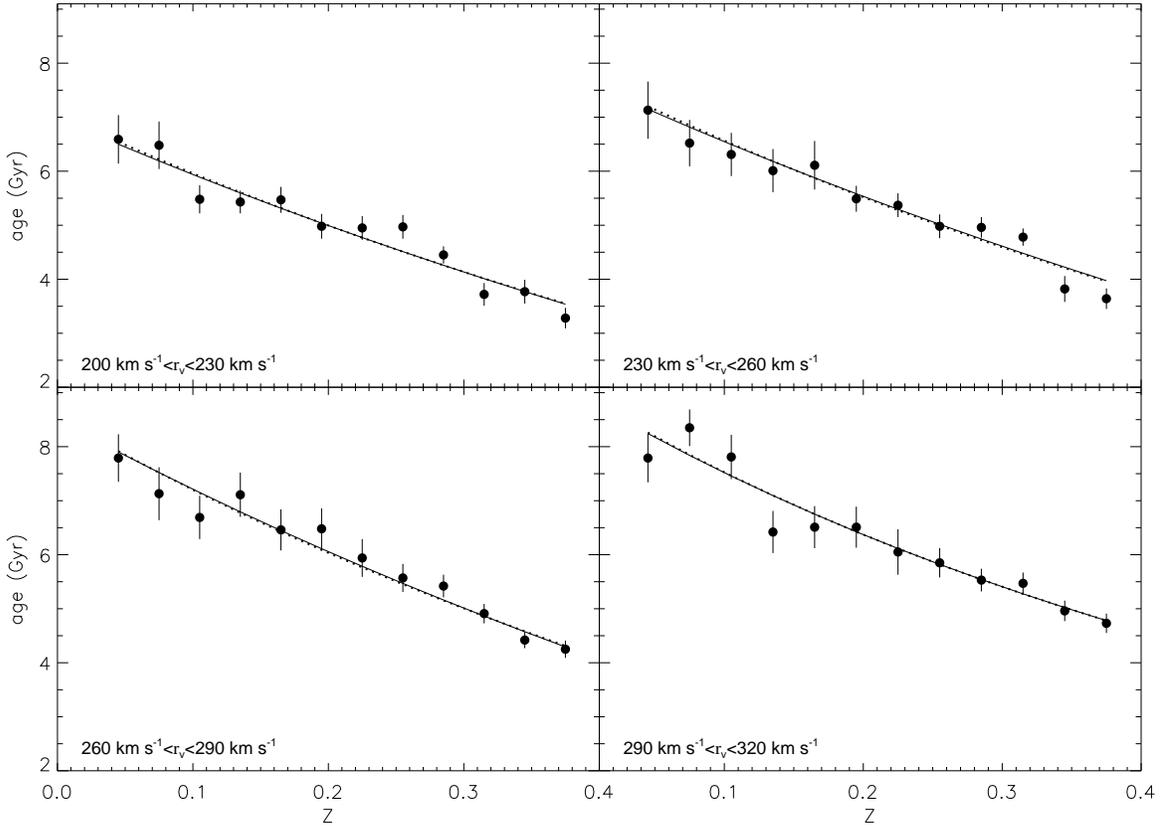}
\caption{ 
The  best fits to the age--redshift relations obtained from each sub-sample
(points with errorbars) by assuming  $H(z)=H_0+H' z$ (dashed lines) and 
a spatially flat $\Lambda$CDM model (solid lines), respectively. The dashed
lines almost overlap the solid lines.
}
\label{fig:f5}
\end{figure*}

\begin{figure*} 
\centering
\includegraphics[scale=1, width=16.0cm]{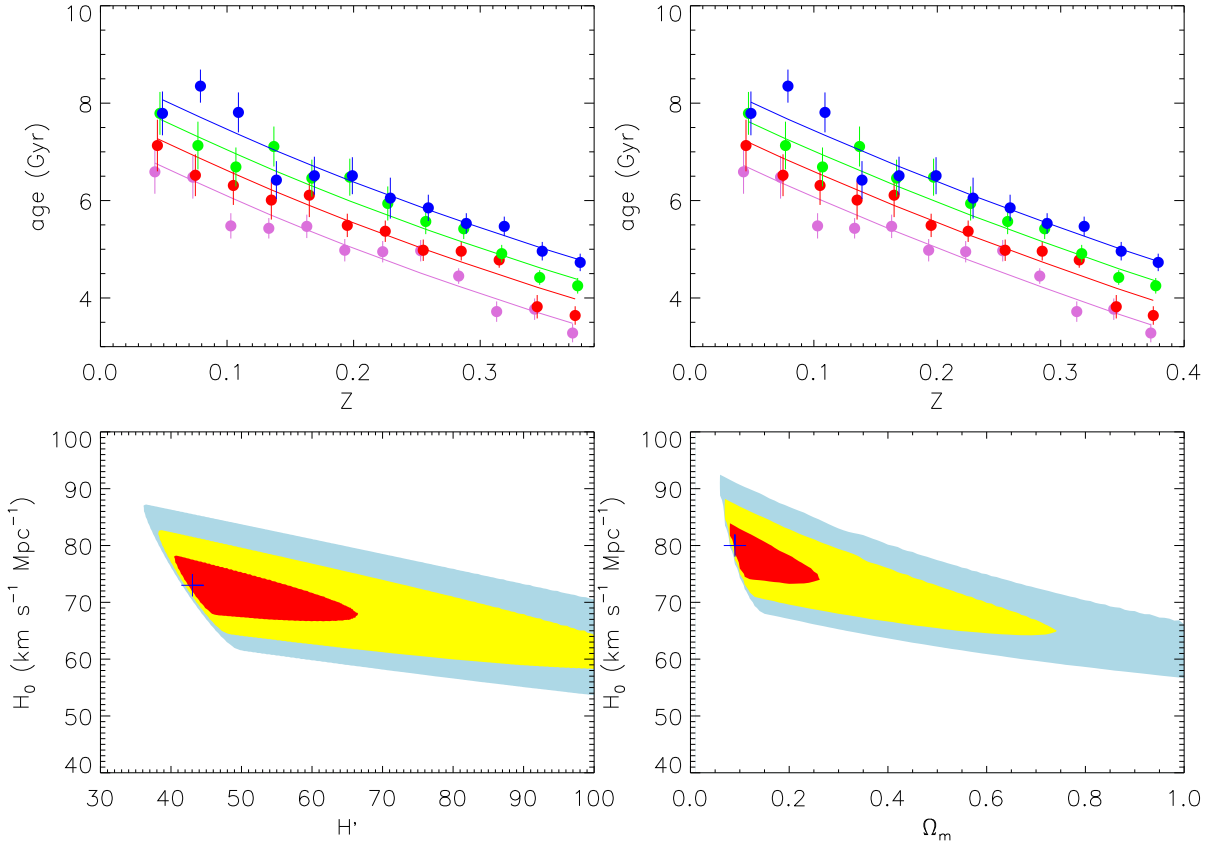}
\caption{ 
The best fit to the age--redshift relations by fitting these relations
obtained from the four sub-samples simultaneously. Left panels show
the results obtained by assuming $H(z)=H_0+H'z$, while right panels
show the results obtained by assuming a spatially flat $\Lambda$CDM model.
The top panels show the best fits (lines) to the age--redshift
relations (points with errorbars) obtained from the sub-sample I (purple),
sub-sample II (red), sub-sample III (green) and sub-sample IV (blue),
respectively. The left-bottom panel show the
confidence levels of the fitting parameters  ($H_0$,$H'$)  by
marginalizing over the formation time $t_{\form1}$, $t_{\form2}$, $t_{\form3}$,
$t_{\form4}$, and the right bottom panel shows the confidence levels of the
fitting parameters ($H_0$,$\Omega_{\rm m}$) by marginalizing over $t_{\form1}$,
$t_{\form2}$, $t_{\form3}$, $t_{\form4}$. The crosses denote the positions of
the best fits, the contours correspond to the $1\sigma$, $2\sigma$, $3\sigma$
levels, respectively. 
}
\label{fig:f6}
\end{figure*}

In order to obtain a model-independent measurement of $H_0$, however, we first
simply assume $H(z)=H_0+H' z$ (but it may be a good approximation only at
low redshift), then the age of the universe is given by 
\begin{equation}
t_{\Univ}=\int_{z}^{\infty} \frac{dz}{(1+z) (H_0+H' z)}.
\end{equation}
Using the standard $\chi^2$ minimization, we fit the age-redshift relation
obtained from each sub-sample to get the best fit of $H_0$ and $t_{\form}$, and
the uncertainty of $H_0$ by marginalizing over $H'$ and $t_{\form}$ (and the
uncertainty of $t_{\form}$ by marginalizing over $H'$ and
$H_0$). Figure~\ref{fig:f5} shows the best fit of the age--redshift relation
for each sub-sample, respectively. The best fits of the Hubble parameter $H_0$
range from $84^{+7}_{-9}\kmsmpc$, $77^{+9}_{-7}\kmsmpc$, $68^{+5}_{-7} \kmsmpc$
to $63^{+7}_{-4}\kmsmpc$ for the sub-samples with velocity dispersions from low
to high (see Table~\ref{tab:t4}), but the mean galaxy formation time can not be
well constrained. 
The age--redshift relations obtained from all the four
sub-samples can also be fitted simultaneously, and only six parameters are now
involved in the fitting, i.e., $H_0$, $H'$, and the mean formation time of the
galaxies in each sub-sample, denoted as $t_{\form1}$, $t_{\form2}$,
$t_{\form3}$, and $t_{\form4}$, respectively, since $H_0$ and  $H'$ are the
same for all the age--redshift relations. By marginalizing over parameters
$H'$, $t_{\form1}$, $t_{\form2}$, $t_{\form3}$ and $t_{\form4}$, we obtain the
best-fit of $H_0=73^{+4}_{-3}\kmsmpc$ (see the left panels of Figure~\ref{fig:f6}) 

\begin{deluxetable}{cccccc}
\tabletypesize{\scriptsize}
\tablecaption{The best fit of the Hubble parameter at the present time $H_0$ }
\tablewidth{0pt} 
\tablehead{ 
\multirow{2}{*}{Sample} & \multicolumn{2}{c}{$H(z)=H_0+H'z$} & \colhead{} &
\multicolumn{2}{c}{Flat $\Lambda$CDM} \\
\cline{2-3} \cline{5-6} 
  & \colhead{$H_0$} & \colhead{$\chi^2_{\nu}$} & \colhead{} & \colhead{$H_0$}  & \colhead{$\chi^2_{\nu}$}
}
\startdata
Sub-sample I      & $84^{+7}_{-9}$  & 1.47 & & $89^{+7}_{-9}$ & 1.43 \\
Sub-sample II     & $77^{+9}_{-7}$  & 1.26 & & $83^{+9}_{-8}$ & 1.21 \\
Sub-sample III    & $68^{+5}_{-7}$  & 0.72 & & $72^{+6}_{-7}$ & 0.66 \\
Sub-sample IV     & $63^{+7}_{-4}$  & 0.86 & & $65^{+7}_{-3}$ & 0.86 \\
Sub-sample III+IV & $68^{+4}_{-5}$  & 0.82 & & $74^{+5}_{-4}$ & 0.78 \\
All sub-samples   & $73^{+4}_{-3}$  & 1.14 & & $80^{+2}_{-4}$ & 1.08 
\enddata
\label{tab:t4}
\tablecomments{Here $H_0$ is in unit of $\kmsmpc$, $\chi^2_{\nu}$
is the reduced $\chi^2$. Columns 2 and 3 list the best-fit value of $H_0$,
its $1\sigma$ error and the reduced $\chi^2$, by assuming
$H(z)=H_0+H'z$; while columns 4 and 5 list the best-fit value of $H_0$,
its $1\sigma$ error and the reduced $\chi^2$ by assuming a spatially flat $\Lambda$CDM model.}
\end{deluxetable}

Assuming spatial flatness for simplicity, we further adopt
equation~\ref{eq:tage} and equation~\ref{eq:tuniv} (given by the standard
$\Lambda$CDM model) to fit the age--redshift relation obtained from each
sub-sample, separately. The best fit of $H_0$ is obtained by marginalizing over
$\Omega_{\rm m}$ and $t_{\form}$ and it ranges from $89^{+7}_{-9}\kmsmpc$,
$83^{+7}_{-8}\kmsmpc$, $72^{+6}_{-7} \kmsmpc$ to $65^{+7}_{-2}\kmsmpc$ for the
four sub-sample velocity dispersion from low to high, respectively (see
Figure~\ref{fig:f5} and Table~\ref{tab:t4}). We also obtain $H_0$ by fitting
the age--relations obtained from all the four sub-sample simultaneously by
marginalizing over $\Omega_{\rm m}$ and the mean formation time of those
galaxies, and the best fit of $H_0$ is $80^{+2}_{-4}\kmsmpc$ (see the right
panels of Figure~\ref{fig:f6}). And the best fit of $\Omega_{\rm m}$ is
$0.09$ and it ranges from $0.08$ to $0.25$, which cannot be constrained
with high accuracy. By marginalizing over other parameters, we obtain
 the best fit of  $t_{\form1}$, $t_{\form2}$, $t_{\form3}$, or $t_{\form4}$
as $9.7^{+0.3}_{-2.7}$~Gyr, $9.2^{+0.3}_{-2.4}$~Gyr, $8.8^{+0.2}_{-2.3}$~Gyr, 
or $8.4^{+0.2}_{-2.2}$~Gyr.  These values of $t_{form}$ seem to be far 
too large than the expectation from the standard $\Lambda CDM$ cosmology for LRGs, 
which are mainly caused by the
poorly constrained $\Omega_{m}$ (=0.09). By assuming a flat universe with
$\Omega_{m}$=0.27 (the concordant cosmological model) and re-fitting the
age-redshift relations for the four velocity dispersion bins simultaneously,
the best fit gives $H_0=73\kmsmpc$, $t_{form1}=5.9$~Gyr, $t_{form2}=
5.4$~Gyr, $t_{form3}=5.0$~Gyr, and $t_{form4}=4.6$~Gyr, respectively.
Clearly the mean formation time of those less 
massive galaxies is smaller than that of those more massive galaxies, which 
is fully consistent with the ``downsizing'' evolution nature of galaxy 
formation. One may also directly obtain the formation time from the 
age--redshift relations by adopting the concordant cosmological model, i.e., 
fixing the Hubble constant $H_0=71 \kmsmpc$, the matter density $\Omega_{\rm m}=
0.27$ and dark energy density $\Omega_{\Lambda}=0.73$, the formation time 
from the age--redshift relations are $t_{\form1}=6.2$~Gyr, $t_{\form2}= 5.7$
~Gyr, $t_{\form3}=5.3$~Gyr, and $t_{\form4}=4.9$~Gyr, respectively. The mean
galaxy formation time is about $5.5$~Gyr for the four velocity dispersion bins, 
which is adopted in Figure~\ref{fig:f3} (represented by the dashed line). 
Note that the obtained $t_{\form}$ is the average age of a population of 
galaxies but neither the age of individual galaxies nor the oldest population 
of stars in those galaxies. Note also there is strong degeneracy between 
$t_{\form}$ and ($H_0$,$\Omega_{\rm m}$) obtained from the fitting, which introduces 
a large uncertainty in the estimation of $t_{\form}$. 

According to the above fittings, obviously a strong constraint on $H_0$ can
still be obtained by either assuming a simple model independent form of the
evolution of $H(z)$ or a flat universe, i.e., $\Omega_{\rm m}+\Omega_{\Lambda}
=1$, although the age--redshift relations are obtained in a limited redshift
range and the uncertainties in the age estimates may be substantial. The $H_0$
estimated from the sub-sample with lower velocity dispersion tends to be higher
than that from the sub-sample with higher velocity dispersion (more massive and
luminous LRGs), which may be due to some bias introduced by the systematical
difference in the assembly history of less massive galaxies and massive
galaxies. \citet{Bro07} pointed out that the evolution of galaxies in the red
sequence is heavily dependent on luminosity, i.e., the lower the luminosity of
the galaxies, the more significant the population of new stars formed since
$z=1$. So the contamination from the population of stars formed at low redshift
(e.g., $z\la 0.4$) is probably more significant in the sub-sample I and
sub-sample II than that for the sub-sample III and sub-sample IV. And the
age--redshift relation estimated from the lower velocity dispersion sub-sample
tends to be shallower than that from the higher velocity dispersion sub-sample,
which may lead to an overestimation of the Hubble parameter up to $\sim 20\%$
(see discussions in Section~\ref{sect:discussion2}). 

As the possible systematic bias may be not significant for the two subs-samples
with high velocity dispersions, we also fit the age--redshift relations
obtained from the two sub-samples simultaneously by assuming either a
spatially flat $\Lambda$CDM model or $H(z)=H_0+H'z$. The best fit of $H_0$ is
either $74^{+5}_{-4}\kmsmpc$ or $68^{+4}_{-5}\kmsmpc$. And the best fit of
$\Omega_{\rm m}$ is $0.07^{+0.28}_{-0.01}$ if assuming a spatially flat
$\Lambda$CDM model.
 
Considerable progress has been made in determining the Hubble parameter over the
past two decades by using many different techniques \citep[e.g.,][]{Fre10}. For
example, the Hubble parameter is estimated to be $H_0=73 \pm 5\kmsmpc$ by using
the tip of the red giant branch as an alternate calibration to the Cepheid
distance scale \citep{Mou08}; $72\pm 4(random) \pm 11 (systematic) \kmsmpc$ by
using the surface brightness fluctuation to determine cosmic distances
\citep{Bla02}; $74.3\pm 3.6 \kmsmpc$ by using Type Ia supernovae \citep{Rie09};
$76.9^{+3.9+10}_{-3.4-8.0} \kmsmpc$ by using the Sunyaev-Zel’dovich effect
\citep{Bon06}; and $68 \pm 2 \kmsmpc$ by using the BAO signature in the matter
power spectrum \citep{Per10}. The Hubble Space Telescope key project yielded a
consistent value of $H_0=72\pm 3 (random)\pm7 (systematic) \kmsmpc$ by
combining the data obtained from different techniques \citep{Fre10}.
\citet{Kom09} also obtained a value of $H_0=70.5\pm 1.3 \kmsmpc$ by combining
the WMAP-5 data with the SNe Ia and BAO data, while \citet{Tam08}
found consistently low values of $H_0$, from several different tracers they
obtain a mean value of $H_0=62.3\pm 1.3 \kmsmpc$. Considering of the possible
systematic error in the $H_0$ estimation by using the age--redshift relation,
our estimates of $H_0$ are fully consistent with those listed above.

\section{Discussion}\label{sect:discussion} 

In this paper, we have obtained the mean age, metallicity and velocity
dispersion for a sample of quiescent galaxies selected from the SDSS DR7 in
different redshift bins.  The age-redshift relation derived from those
quiescent galaxies is consistent with the expectation from the $\Lambda$CDM
cosmology and may provide a good estimate of the Hubble parameter ($H_0$).
However, the estimate of $H_0$ is valid only if those quiescent galaxies are
passively evolving and the galaxies in different redshift bins represent the
same population formed more or less at the same time. In order to check whether
these requirements are satisfied for the quiescent galaxy sample selected in
this paper, we perform some tests below to investigate the evolution effects
due to different star formation history or galaxy mergers, and illustrate that
those quiescent galaxies are indeed more or less formed at the same epoch using
the evolution of their average colors.

\subsection{Star formation history}
\label{sect:discussion1} 

According to the selection criteria, most galaxies in our sample
should be quiescent galaxies and supposed to be passively evolving.
However, the star formation history of those galaxies may not be a
single burst. To test whether there are significant younger stellar
populations in the galaxies, we re-do the full spectrum fitting for
each combined spectrum by adopting two stellar components, one young
stellar population (YSP) and one old stellar population (OSP). The age
of the young stellar population is assumed to be in the range from
$0.1$~Gyr to $1$~Gyr, while for the old population it is assumed to be
in the range from $1$~Gyr to $15$~Gyr. For both populations, the
metallicity is a free parameter without restrictions.  According to
the fitting results, we note here that the fitting age of the YSP
always reaches the edge of its limits (the age of YSP is always either
$0.1$~Gyr or $1$~Gyr), which means that a YSP with an age in the given
range can not be found. Furthermore, even if there exist a YSP, the
light fraction (LF) is very small ($2.8\%$ on average) so that it can
be neglected. Therefore, we conclude that the YSP (the age of which
$\textless 1$~Gyr) is negligible and not required in the fitting. We
also test the cases by setting a larger upper limit on the age of the
YSP, e.g., $2$~Gyr, $3$~Gyr or even $4$~Gyr (or alternatively a lower
limit on the age of the old stellar population, i.e., $4$~Gyr to
$6$~Gyr), and also find that no significant YSP is required by the
fitting. All these tests suggest that most of the quiescent LRGs may
be passively evolving and not experience significant recent ($\la
2$~Gyr) star formation. For those galaxies at low redshift bins,
however, we find there may exist YSP's with age $\sim 3$~Gyr, possibly
due to the later major mergers (see discussions in Section
~\ref{sect:discussion2}).

 To close this sub-section, we note here that
\citet{Toj10} find bright LRGs being consistent with pure passive
evolution while faint LRGs slightly deviating from pure passive
evolution as revealed by the evolution of the number and luminosity
density of LRGs as well as that of their clustering. The lesser
passiveness of LRGs with smaller velocity dispersion seems not to be
able to be directly revealed by the combined spectra of LRGs studied
in this paper, which might be due to that the signature of YSPs
(probably with quite different ages) in (some of) the LRGs may be
diluted or smoothed due to the co-adding of a large number of LRG
spectra in our analysis. High S/N spectra of individual LRGs may be
helpful to clarify the less passiveness of small LRGs, however, the
S/N of most LRGs in our sample are only slightly larger than 10 and
not sufficiently high for clarifying this problem.  

\subsection{Combined spectrum vs. single spectrum}
\label{sect:discussion0} 

\begin{figure} 
\centering
\includegraphics[scale=1.0,width=8.0cm]{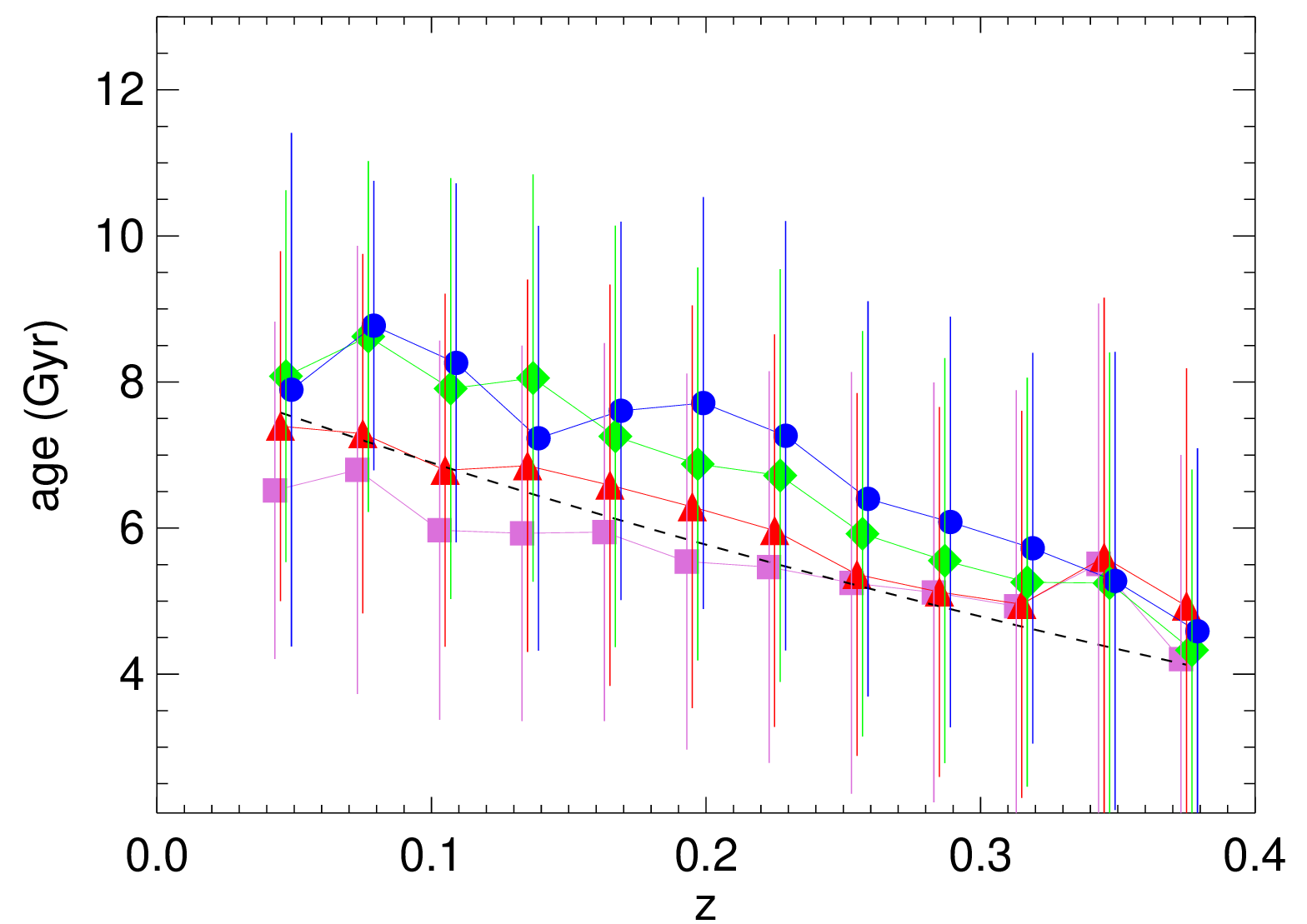}
\caption{
The mean age-redshift relation derived by fitting each single spectrum in the
LRG sample with \texttt{ULySS} for each redshift bin of each of the four
sub-samples, $200\kms<\sigma_{v} \le230\kms$ (purple squares),
$230\kms<\sigma_{v}\le260\kms$ (red triangles), $260\kms<\sigma_{v}\le290\kms$
(green diamonds), and $290\kms<\sigma_{v}\le320\kms$ (blue circles),
respectively.  The dashed line indicates $t_{\Univ}(z)-5.5\,\textnormal{Gyr}$ ,
where $t_{\Univ}(z)$ is the age of the universe for a $\Lambda$CDM cosmology
and is for reference only.
}
\label{fig:f7}
\end{figure}

The combined spectrum in each redshift and velocity bin may only represent the
mean spectrum of galaxies in that bin. Does the physical properties derived
from the model fitting of this combined spectrum represent the mean properties
of all galaxies in that bin?  In order to test this, we fit each single
spectrum for all the galaxies in each velocity dispersion bin and redshift bin
with \texttt{ULySS}. The initial settings are the same as that in
Section~\ref{sect:combined spectrum}.  After obtaining the best fit of the age
for each single spectrum, we calculate the mean of the ages of those galaxies
which belong to the same velocity dispersion and redshift bin and its standard
deviation.  Figure~\ref{fig:f7} shows the mean age-redshift relation for every
single spectrum. As seen from Figure~\ref{fig:f7}, the mean age-redshift
relation obtained from the single spectrum fitting is still consistent with
that obtained from the combined spectra but with much larger errors. 

\subsection{Major mergers of quiescent galaxies}
\label{sect:discussion2} 

The differences among the model spectra become small if the ages of the stellar
populations are larger than $4$~Gyr. Therefore, the uncertainties in the
estimates of the ages and metallicities of galaxies with low $S/N$ spectra are
substantial. For this reason, we have combined the spectra of galaxies in each
redshift and velocity dispersion bin together to improve the $S/N$. However,
the age difference obtained by modeling the combined spectra can represent the
age difference of the universe only if those galaxies at different redshifts are
the same population and were formed more or less in the same epoch. There is a
potential caveat to this approach, i.e., if a significant fraction of quiescent
galaxies experience major mergers and thus some star formation at low redshift
$z\la 0.4$, then some quiescent galaxies in the redshift bin $0.03-0.06$ were
formed at $z<0.4$ and they were not represented by those galaxies in the
redshift bin $0.36-0.39$. It is important to check the effect of the major
merger of galaxies on the age-redshift relation.

The major merger rate of galaxies is defined as the number of mergers per
galaxy with mass larger than a threshold ($M_{\ast}$) per unit time and denoted
as $dN_{\rm mrg}/dt$, and it can be roughly estimated by the fitting formula
given by \citet{Hop10}, i.e.,  
\begin{equation}
\frac{{\rm d}N_{\rm mrg}}{{\rm d}t} = A(M_{\rm
min})(1+z)^{\beta(M_{\rm min})}[{\rm per\ galaxy}],
\end{equation}
where the normalization is
$$A(M_{\rm min})_{\rm major} \approx 0.02[1 + (M_{\rm min}/M_{0})^{0.5}] {\rm
Gyr^{-1}},$$ 
and the redshift evolution is 
$$\beta(M_{\rm min})_{\rm major}
\approx 1.65 - 0.15\log(M_{\rm min}/M_{0}),$$ 
and $M_{0}\equiv 2\times10^{10}\,{M_{\sun}}$.  
For the galaxies in our sample, they are
generally brighter than $3L^{\ast}$ \citep{Eis01} and have stellar masses in
the range from $10^{11} {M_{\sun}}$ to a few times  $10^{12} {M_{\sun}}$
according to \citet{Tal11}.  Assuming that the minimum mass of those galaxies
is $M_{\rm min}=10^{11} {M_{\sun}}$, the average number of major mergers
experienced by a galaxy at redshift $z$ since $z=0.40$ is 
\begin{equation}
N_{\rm mrg}(z)=\int_{z}^{0.40} \frac{dN_{\rm mrg}}{dt}  |\frac{dt}{dz}| dz,
\end{equation}
where $dt/dz$ is given by Eq.~(\ref{eq:dtdz}).  The fraction of galaxies in a
redshift bin $z\pm dz$ which experienced major mergers since $z=0.4$ is $\sim
N_{\rm mrg}(z)$, and $N_{\rm mrg}(z)\sim 0.33$, $0.26$, and $0.01$ at $z=0.03$,
$0.11$, and $0.39$ for our sample, respectively. Note that almost all the major
mergers are dry mergers for massive galaxies with mass $>10^{11}M_{\odot}$
similar to that in our sample \citep{Hop10}. These dry mergers may also lead to
new star formation or star burst in the galactic centers, and thus introduce a
systematic bias to the age--redshift relation, because the mean age obtained
from the combined spectra by the GS model at low redshift bins may be
systematically underestimated due to the additional population of stars formed
later. 

We check the effect of these major mergers on the age--redshift
relation as following. First, we assume an underlying age-redshift
relation according to the $\Lambda$CDM, i.e., the age of the galaxy at
redshift $z$ is $t_{\Univ}(z)-t_{\form}$, and $H_0=71\kmsmpc$,
$\Omega_{\rm m}=0.27$, $t_{\form}=5.5~{\rm Gyr}$ are assumed.  Here
$t_{\form}=5.5~{\rm Gyr}$ is adopted according to the average
formation time obtained by fitting the age-redshift relation for four
velocity dispersion bins (see Section ~\ref{sect:Hubble_parameter} for
details).  However, Thomas et al. (2005) found that vigorous star
formation episodes in massive galaxies ofter occur at $z\sim 2-5$,
corresponding to a cosmic age of less than $3.3$~Gyr.  The $t_{form}$
obtained from the age--redshift relations for the quiescent LRGs in
this paper is larger than that obtained by Thomas et al. (2005),
which needs further investigation. 
As the galaxies in our sample are all
LRGs,  the age of the newly formed stellar population by major mergers, if
significant, is at least $1$~Gyr, which corresponds to $\delta_{\rm z}\sim
0.1$, as the migration of galaxies driven by mergers from blue cloud to red
sequence may last $\sim1$~Gyr \citep[see][]{Sch10}. If a galaxy at redshift $z$
is the remnant of a major merger, the age of the younger stellar population in
it is approximately $1/2[t_{\Univ}(z)-t_{\rm U}(z=0.4)]+1{\rm Gyr}$. Then a
``forged'' combined spectrum of galaxies at redshift bin z can be approximately
generated by two stellar populations, i.e., an old stellar population with the
age $t_{\Univ}(z)-t_{\rm form}$, and a young stellar population with age
$\sim1/2[t_{\Univ}(z)-t_{\Univ}(z=0.4)]+1{\rm Gyr}$. The fraction of major
mergers is $$N_{\rm mrg}({z+\delta_{\rm z}}) = \int_{z+{ \delta_{\rm
z}}}^{0.40+ \delta_{\rm z}} \frac{dN_{\rm mrg}}{dt } \left|\frac{dt}{dz}\right|
dz.$$ \citet{Hop10} have shown that the fraction of the young population
generated by major mergers is $\sim 0.05-0.10$ in galaxies with
$>10^{11}M_{\odot}$ (see their Figure~14).  Here we assume it is $\sim 0.07$.
Note that this fraction may be an upper limit as the LRGs studied here are
quiescent ones, in which the star formation generated by major mergers might be
even less. Therefore, the fraction of the young population contributing to the
combined spectrum is $\sim 0.07\times N_{\rm mrg}(z)$, and the fraction of the
old population is $1-0.07\times N_{\rm mrg}(z)$. We fit these spectra obtained
for each redshift by the same method as that in Section~\ref{sect:combined
spectrum} and obtain the age for these combined spectra. Similar to that in
Section~\ref{sect:Hubble_parameter}, we obtain the best fit of
$H_0=86^{+2}_{-4}\kmsmpc$ by fixing $\Omega_{\rm m}=0.27$.

From the above calculation, we conclude that $H_0$ may be systematically
overestimated by up to $\sim 20\%$ if using the age-redshift relation obtained
from the combined spectra. Considering this systematic bias, $H_0$ estimated
from the age--redshift relation in this paper is consistent with those
estimated by other techniques. Note also that the systematic bias introduced
to the $H_0$ estimate appears not significant for the two sub-samples with the
highest velocity dispersions, which might mean that those very massive LRGs are
really quiescent and have approximately zero star formation at redshift $z\la
0.4$. \citet{Toj10} have pointed out that the brightest galaxies show the
smallest departure from pure passive evolution. Therefore, the most massive
LRGs with velocity dispersion $\ga 300\kms$ may be efficient tools to constrain
the cosmological parameters, such as $H_0$, through the age--redshift relations
extracted from their spectra.

We also check whether the g-r color of the galaxies having the combined spectra
in each redshift bin is consistent with that of the model galaxies that have
the same spectrum as the combined spectrum in the highest redshift bin. To do
this, we extract a model spectrum from the GS model grid, whose age and
metallicity are the best fitting results with \texttt{ULySS} for redshift bin
0.36-0.39, then we let this spectrum evolve toward the low redshift. That is,
if the age and metallicity of the combined spectrum are t and $Z_{\rm Fe/H}$ at
z=0.375, then at a low redshift z, we extract a spectrum from model grid whose
age is $t+\delta t$ and let the metallicity is fixed to $Z_{\rm Fe/H}$, where
$\delta t=t_{\rm U}(z)-t_{\Univ}(z=0.375)$. We repeat this process till
z=0.045, and denote the spectra obtained from the above processes as the
``forged spectra".  For each combined spectrum there is also a model spectrum
to fit it.  We denote this spectrum as the ``model spectrum". Then we calculate
the g-r color with SDSS filter of those ``forged spectra" and ``model spectra"
with \texttt{IRAF} task \texttt{SYNPHOT}. We find that the g-r color of the
``model spectra" are almost the same as that of the ``forged spectra", which
means that the merger effect is not significant for our sample.  Furthermore,
we pick out those spectrum whose $S/N > 30$ from our sample as a sub-sample to
derive their physical property with full spectrum fitting.  This sub-sample has
1386 galaxies, and these are almost all concentrated in the redshift range from
0.02 to 0.2.  We fit these galaxies with SSPs. The initial settings are the
same as that in Section~\ref{sect:combined spectrum}. Similarly, we extract the
corresponding spectrum from the GS model according to the best fit. Then we
calculate the g-r color and find that dispersion of the g-r color  is very
small compared with that of the combined spectrum, which means the results of
the single spectrum are consistent with that of the combined spectrum.  

\subsection{Model dependence}
\label{sect:discussion3} 

In this paper, we use the GalexEV/SteLib model to fit the combined spectra, as
its wavelength coverage is wider compared with the other models provided by the
\texttt{ULySS} package. For completeness, we also test the Pegase-HR/Elodie3.1
model and the Vazdekis/Miles model, and find there is some model dependence.
The model dependence may be caused by the limitation of wavelength coverage as
the Pegase-HR/Elodie3.1 model covers the wavelength from $3900$\AA$-6800$\AA,
and for the Vazdekis/Miles model it is $3540$\AA$-7409$\AA. This result is
consistent with that of \citet{Ver05}, in which they analyzed the photometric
data of a large sample of elliptical galaxies and found that $H_0=72\pm~10
\kmsmpc$ and $53\pm10~\kmsmpc$ for two different stellar population synthesis
models, i.e., PEGASE and GISSEL , respectively.
The main
ingredients of those models are the stellar evolution tracks, the stellar
library, the IMF, the grids of ages and metallicities, and the SFHs, etc. Each
model may have different settings in one or more of those ingredients. As
demonstrated by \citet{Chen10}, at present there are still some differences in
the output between these different models.  The model dependence may be
avoidable if carefully choosing a compatible stellar population synthesis model
for the problem to be studied. 
We note here that is also interesting to test the model given by
\citet{Mar11}, which is based on the fuel consumption theorem and quite
different from the other models. However, it is not easy to test this model
because it is not yet included in the  \texttt{ULySS} code. 

From the above tests, we conclude that the age--redshift relation obtained
from the combined spectra of a large sample of quiescent LRGs can be used
to constrain the Hubble parameter $H_0$ (and possibly other cosmological
parameters if combining with other data sets). If a large sample of very
massive quiescent LRGs can be obtained by the future surveys, such as the
Baryon Oscillation Spectroscopic Survey (BOSS),
 which is less affected by the systematic bias due to new star formation
at low redshift, the $H_0$ may be able to be determined with substantial 
accuracy.

\section{Conclusions}\label{sect:conclusion}

In this paper, we selected $23,883$ quiescent LRGs from the SDSS DR7 in the
redshift range from $0.03$ to $0.39$, by setting a threshold of zero emission
(at a $2-\sigma$-level) of the H$_\alpha$ and [O{\sc{ii}}] lines directly
obtained from the MPA/JHU catalogue. The quiescent LRG sample is divided into
four sub-samples according to galaxy velocity dispersions. For each sub-sample,
the spectra of galaxies in each of the $12$ redshift bins (from $z=0.03$ to
$0.39$ with a step of $\delta z=0.03$) are combined together to obtain a high
$S/N$ combined spectrum. Using the full spectrum fitting method, the
luminosity-weighted physical properties, such as the velocity dispersion, the
metallicity and the age, of those quiescent LRGs are obtained from the combined
spectra by adopting a single population synthesis model, i.e., the
GalexEV/SteLib model. Using Monte-Carlo simulations,
we find that the model results are robust and reliable. We argue that the
age--redshift relation estimated from the LRG sample could be systematic biased
because of the contamination from a possible younger stellar population formed
at $z\la 0.4$ as consequence of major mergers. This bias is most
significant for LRGs with smaller velocity dispersions but insignificant for
the most massive LRGs. Considering of this systematic bias, the age--redshift
relation obtained from the model fittings is fully consistent with the
expectations from the $\Lambda$CDM cosmology.  

The Hubble parameter $H_0$ is first estimated by using the age--redshift
relation obtained from each sub-sample, and its value ranges from $89^{+7}_{-9}
\kmsmpc$, $83^{+9}_{-8}\kmsmpc$, $72^{+6}_{-7}\kmsmpc$, to $65^{+7}_{-3}\kmsmpc$ for the four sub-samples with velocity
dispersions from low to high, respectively. The large value of the $H_0$ estimated from the
sub-samples with low velocity dispersion is probably due to the systematic
bias, which can be as high as $\sim 20\%$. 
Using the age--redshift relations obtained from the two sub-samples with
high velocity dispersions or the sub-sample with the largest velocity
dispersion, we find $H_0= 74^{+5}_{-4}\kmsmpc$ or $H_0=
65^{+7}_{-3}\kmsmpc$ if assuming a spatially flat $\Lambda$CDM
cosmology, which may be less affected by the systematic bias, close to the
true $H_0$, and are
well consistent with the best estimates through other techniques.
However, it needs further test on whether those most massive galaxies
are truly passively evolving or not.

In summary, we have demonstrated that the age--redshift relation of quiescent
galaxies can be reliably estimated by using the full spectral fitting method if
the $S/N$ of their spectra are sufficiently high. We conclude that some
cosmological parameters, such as the Hubble parameter, can be constrained with
considerable accuracy through the age--redshift relation obtained from those
most massive LRGs, which is totally independent of other methods. With future
surveys like BOSS, the Hubble parameter may be tightly constrained
by the age--redshift relation obtained from the most massive quiescent LRGs.

\acknowledgments

We thank the anonymous referee for helpful comments. We are grateful
to Raul Jimenez for conversations on constraining $H_0$ through the
age--redshift relation and to Licia Verde for useful comments on the
paper.  Gaochao Liu thanks Mina Koleva for helpful discussions on
using  \texttt{ULySS} and thanks Yanchun Liang, Yingchun Wei,
 Yan Gong, Haijun Tian, Xiaoyan Chen for their kindly help.
 We thank the NSFC grant support under No.~11003022,
10973017, 11033001, 11073024, and the CAS grant KJCX2-EW-W01.

Funding for the SDSS and SDSS-II has been provided by the Alfred P. Sloan
Foundation, the Participating Institutions, the National Science Foundation,
the U.S. Department of Energy, the National Aeronautics and Space
Administration, the Japanese Monbukagakusho, the Max Planck Society, and the
Higher Education Funding Council for England. The SDSS Web Site is
http://www.sdss.org/.

The SDSS is managed by the Astrophysical Research Consortium for the
Participating Institutions. The Participating Institutions are the American
Museum of Natural History, Astrophysical Institute Potsdam, University of
Basel, University of Cambridge, Case Western Reserve University, University of
Chicago, Drexel University, Fermilab, the Institute for Advanced Study, the
Japan Participation Group, Johns Hopkins University, the Joint Institute for
Nuclear Astrophysics, the Kavli Institute for Particle Astrophysics and
Cosmology, the Korean Scientist Group, the Chinese Academy of Sciences
(LAMOST), Los Alamos National Laboratory, the Max-Planck-Institute for
Astronomy (MPIA), the Max-Planck-Institute for Astrophysics (MPA), New Mexico
State University, Ohio State University, University of Pittsburgh, University
of Portsmouth, Princeton University, the United States Naval Observatory, and
the University of Washington.

\end{document}